\documentclass[prd,twocolumn,showpacs,floatfix]{revtex4}
\usepackage{hyperref,amssymb,amsmath,mathrsfs,bm,graphicx}

\begin{document}
\title{Maxwell's  Demon and the Problem of Observers in General Relativity}
\author{L. Herrera}
\email{lherrera@usal.es}
\affiliation{Instituto Universitario de F\'isica
Fundamental y Matem\'aticas,  Universidad de Salamanca, Salamanca 37007, Spain}

\begin{abstract}
The fact that real dissipative (entropy producing) processes may be  detected by non-comoving observers (tilted), in systems that appear to be isentropic for comoving observers, in general relativity,  is explained in terms of the information theory, analogous with the explanation of the Maxwell's demon paradox. 
\end{abstract}
\maketitle
\section{INTRODUCTION}

 Observers play an essential role in quantum mechanics where the very concept of reality is tightly attached to the existence of the observer, as ingeniously illustrated by the well known Schrodinger's cat paradox.  In the  quantum mechanics terminology, we say that the observer produces the collapse of the wave function.

However, it is generally assumed that observers do not play a similar role in classical (non~quantum) theories. However, is this assumption really justified? As we shall see here, the answer to such a question is negative. Indeed, the role of observers in General Relativity is a fundamental one, and reminds one, in some sense, of its role in quantum physics, namely: a complete understanding of some gravitational phenomena requires the inclusion of the observers in the definition of the physical system under consideration.
 
 In order to make our case, let us first recall  that, in relativistic hydrodynamics, different observers assign different four-velocities to a given fluid distribution. This simple fact is at the origin of  an ambiguity in the description of the source of the gravitational field (whenever it is represented by a fluid distribution). 
 
Thus, one may face the situation when one of the congruences corresponds to comoving observers, whereas the other is obtained by applying a Lorentz boost to the comoving observer's  frame (this Lorentz boosted congruence is usually referred to as the tilted congruence). 

The strange fact then appears that systems that are isentropic for comoving observers may become dissipative for tilted observers (see \cite{4,5,7,9,38,2S,t3,t3B,t4,t5} and references therein).

We shall illustrate this situation with some examples, and shall provide an explanation based on the theory of information. More specifically, we shall see that an argument similar to the one put forward by Bennet  \cite{ben} to solve the Maxwell's demon paradox \cite{max} may be used to explain the very different pictures of a given system, presented by different congruences of observers  in general~relativity.

 As we shall see, the essential fact is that, when we pass from comoving observers (which assign zero value to the three-velocity of any fluid element) to tilted observers, for whom the three-velocity represents another degree of freedom, the erasure of the information stored by comoving observers (vanishing three velocity) explains the presence of dissipative processes (gravitational radiation included) detected by tilted observers.

\section{COMOVING AND TILTED OBSERVERS}
In order to grasp the essence of  the problem under consideration, it is important  to understand how the tilted (non-comoving) congruence may be obtained from the comoving one. In what follows, we present the general scheme for doing that.
Thus, let us consider  a  congruence of observers that are  comoving with an arbitrary fluid  distribution; then, the four-velocity for that congruence, in some globally defined coordinate sytem, reads
\begin{equation}
V^\mu =(V^0,0,0,0).
\label{vMc}
\end{equation}

In order to obtain the four-velocity corresponding to the tilted congruence (in the same globally defined coordinate system), one proceeds as follows.

We have first to perform a (locally defined) coordinate transformation to the  Locally Minkowskian Frame (LMF). Denoting  by $L _\mu ^\nu$  the local coordinate transformation matrix and, by $\bar V^\alpha$, the components of the four-velocity in such LMF, we have:
\begin{equation}
\bar V^\mu = L^\mu_\nu V^\nu.
\label{V}
\end{equation}

Next, let us apply  a Lorentz boost to the LMF associated with  ${\bar V}^\alpha$, in order to obtain the (tilted)  LMF with respect to which a fluid element is moving with some, non-vanishing, three-velocity.

Then, the four-velocity in the tilted LMF is defined by:
\begin{equation}
\tilde {\bar V}_{\beta}=\Lambda^\alpha_\beta \bar V_\alpha,
\label{1}
\end{equation}
where  $\Lambda^\alpha_\beta$ denotes the Lorentz matrix.

Finally, we have to perform a  transformation from the tilted LMF, back  to the (global) frame associated with  the line element under consideration. Such a transformation, which obviously only exists locally,  is defined by the inverse of  $L _\mu ^\nu$, and produces the four-velocity of the tilted congruence in our globally defined coordinate system, say $\tilde V^\alpha$.

In the following sections, we shall present several examples of tilted space-times, which illustrate the sharp differences in their interpretations, with respect to the picture obtained by the comoving observers. To avoid any confusion, it must be kept in mind that the global coordinate system in each example is the same  for both congruences of observers (comoving and tilted). In addition, it must be stressed that these congruences are not related by any global coordinate transformation. They are related by   the process described above.

\section{Tilting the  Lemaitre--Tolman--Bondi Congruence}

The Lemaitre--Tolman--Bondi metric (LTB) is an exact solution to Einstein's equation \cite{25, 26, 27, 27bis}, which, as seen by a congruence of comoving observers, describes spherically symmetric distributions of   geodesic, shearing, and vorticity free, inhomogeneous non-dissipative dust.
The magnetic part of the Weyl tensor vanishes, whereas its electric part may be defined through a single  scalar function.
If we put the shear or the the Weyl tensor equal to zero, the LTB spacetime becomes the Friedman--Robertson--Walker spacetime.

The general form of LTB metric is defined by:
\begin{equation}
ds^2=-dt^2+B^2dr^2+R^2(d\theta^2+\sin^2\theta d\phi^2),
\label{1}
\end{equation}
where $B(r,t)$ and $R(r,t)$ are functions of their arguments, and as a consequence of the Einstein~equations

\begin{equation}
B(t,r)=\frac{R^\prime}{\left[1+k(r)\right]^{1/2}},\label{BTB}
\end{equation}
where $k$ is an arbitrary function of $r$ and prime denotes derivative with respect to $r$.

The energy momentum tensor describing a dust distribution with energy density $ \mu$ in comoving coordinates takes the usual form:

\begin{equation}
T_{\mu\nu}= \mu V_\mu V_\nu.
\label{Tpo}
\end{equation}
Obviously, for the comoving observer, the fluid is geodesic.

However, if we now tilt the comoving observer, then, as it has been shown, the spacetime  appears   to be sourced by a dissipative anisotropic fluid distribution, and furthermore the fluid is no longer geodesic \cite{38}. 
The important point to stress here is that the tilted observer detects a real dissipative process (entropy producing) as it follows from the discussion on the generalized Gibbs equation (see~\cite{38} for details).

Obviously, due to the spherical symmetry, the magnetic part of the Weyl tensor also vanishes for the tilted observer, implying that no gravitational radiation is detected by the latter.

\section{Tilting the Szekeres Congruence}
In the example analyzed in the previous section, the fluid distribution was spherically symmetric, thus it is interesting to wonder what happens when we consider fluid distributions non restricted  by this symmetry. For doing so, let us consider the Szekeres spacetime \cite{s1, s2}.

Indeed, Szekeres dust models have no Killing vectors and therefore represent an interesting generalization of LTB spacetimes. When analyzed from the point of view of comoving observers, the~Szekeres spacetime is sourced by a geodesic non-dissipative dust, without vorticity. In addition, as in the LTB case, the magnetic part of the Weyl tensor vanishes, implying that there is no gravitational~radiation.

In this case, the line element is given by:
\begin{equation}
ds^2=-dt^2 + \frac{(R^\prime E-R E^\prime)^2}{E^2 (\epsilon + f)} dr^2 + \frac{R^2 }{E^2}(dp^2+dq^2),
\label{sz}
\end{equation}
where a prime denotes a derivative with respect to $r$,  $R=R(t,r)$,  $\epsilon = \pm1,0$ and $f=f(r) > -\epsilon$ is an arbitrary function of $r$. We number the coordinates $x^0=t, x^1=r, x^2= p, x^3=q$.

The function $E$ is given by
\begin{equation}
E(r,p,q)=\frac{S}{2}\left[\left(\frac{p-P}{S}\right)^2 + \left(\frac{q-Q}{S}\right)^2 + \epsilon\right],
\label{E}
\end{equation}
where $S=S(r)$, $P=P(r)$ and $Q=Q(r)$ are arbitrary functions.

From Einstein equations, it follows that  $R$ satisfies the equation

\begin{equation}
\dot R^2=\frac{2M}{R}+f,
\label{1}
\end{equation}
where a dot denotes derivative with respect to $t$, and $M=M(r)$ is an arbitrary function. From the above equation, it follows that

\begin{equation}
\ddot R=\frac{-M}{R^2},
\label{2}
\end{equation}
from where the meaning of $M$ as an effective gravitational  mass becomes evident.

However, the above picture drastically changes when the matter content is analyzed by a tilted~congruence.

Indeed, as shown in \cite{2S}, tilted observers detect a dissipative, anisotropic fluid that is no longer geodesic and furthermore is endowed with vorticity. As for the LTB case, the dissipation detected by tilted observers is ``real'' in the~sense that there is an increasing of entropy. However, even in the~tilted version, the magnetic part of the Weyl tensor vanishes, and so tilted observers do not detect gravitational radiation.

\section{Tilted Shear-Free Axially Symmetric Fluids}
In the examples analyzed in the two previous sections, the Lorentz boost applied to the comoving congruence, in order to obtain the tilted one, was always directed along one of the coordinate axis~($r$). We shall now consider a much more general situation, where the boost is applied along two independent directions.

We shall consider axially symmetric fluids, which for the comoving observer are geodesic, shear-free non-dissipative, and vorticity free.

The line element reads \cite{sg}

\begin{equation}
ds^2=-dt^2+B^2(t)\left[dr^2+r^2d\theta ^2+R^2 (r,\theta)d\phi ^2\right],\label{len}
\end{equation}
where $B(t)$ and $R(r,\theta)$ are  functions of their arguments satisfying the Einstein equations,  and, from regularity conditions at the origin, we must require $R(0,\theta)=0$. 

For the comoving observer, the energy momentum--tensor in the ``canonical'' form reads:
\begin{eqnarray}
{T}_{\alpha\beta}&=& (\mu+P) V_\alpha V_\beta+P g _{\alpha \beta} +\Pi_{\alpha \beta},
\label{6bis}
\end{eqnarray}
where, as usual, $\mu, P,  \Pi_{\alpha \beta}, V_\beta$ denote the energy density, the isotropic pressure, the anisotropic stress tensor and  the four-velocity, respectively.

For the comoving congruence, the anisotropic tensor depends on a single scalar function,   and   the four-velocity vector reads:

\begin{equation}
V^\alpha =\left(1,0,0,0\right); \quad  V_\alpha=\left(-1,0,0,0\right)
\label{m1}
\end{equation}
(see \cite{sg} for details). 

In addition, as shown in \cite{sg}, the magnetic part of the Weyl tensor calculated by means of the four-velocity vector (\ref{m1}) vanishes and the electric part is defined through a unique scalar function.

The above picture is drastically changed when the system is analyzed by a tilted congruence of observers, as {we shall now see
}( see  \cite{hdc} for details). 

For doing so, we have to obtain first the tilted congruence and all the associated kinematical variables, applying the procedure sketched above, for the case when the boost is applied along the $r$ and the $\theta$ directions.

Thus, we obtain for the tilted four-velocity (see \cite{hdc} for details):

\begin{equation}
\tilde V_\alpha=(-\Gamma, B\Gamma v_1, Br \Gamma v_2, 0);\quad \tilde V^\alpha=(\Gamma, \frac{\Gamma v_1}{B}, \frac{ \Gamma v_2}{Br}, 0),
\label{til4}
\end{equation}
where $\Gamma\equiv \frac{1}{\sqrt{1-v^2}}$, $v^2=v^2_{1}+v^2_{2}$, and $v_{1}$, $v_{2}$ are the two non-vanishing  components of the three-velocity of a fluid element as measured by the tilted observer.

We can now calculate all the kinematical variables for the tilted congruence. The result shows  that now the four-acceleration, as well as  the shear and the vorticity are  non vanishing (see \cite{hdc} for details).

In addition, for the tilted congruence, the electric part of the Weyl tensor has  three independent non-vanishing components and the magnetic part of the Weyl tensor is non-vanishing, and defined through  two components. Thus, we may write these two  tensors in terms of five  tetrad components  ($ \tilde{\mathcal{E}}_{I}$, $\tilde{\mathcal{E}}_{II}$, $\tilde{\mathcal{E}}_{KL}$, $\tilde{H}_{1}$, $\tilde{H}_{2}$), respectively.

For the tilted observers, the fluid distribution is described by the energy momentum tensor:

\begin{equation}
\tilde T_{\alpha\beta}= (\tilde \mu_+\tilde P) \tilde V_\alpha \tilde V_\beta+\tilde P g _{\alpha \beta} +\tilde \Pi_{\alpha \beta}+\tilde q_\alpha \tilde V_\beta+\tilde q_\beta \tilde V_\alpha.
\label{6bist}
\end{equation}

It should be noticed that now the system appears to be dissipative, with the heat flux vector defined  through two independent scalar functions $\tilde q_{(1)}, \tilde q_{(2)}$ and the anisotropic tensor is defined by three independent scalars $\tilde \Pi_{I}, \tilde \Pi_{II}, \tilde \Pi_{KL}$.

From the above expressions, we can calculate the super-Poynting vector in terms of only  two  scalar functions
$\tilde P_{(1)}, \tilde P_{(2)}$,
where
\begin{widetext}
\begin{eqnarray}
\tilde P_{(1)}=
\frac{2\tilde{H}_{2}}{3}\left(2\tilde{\mathcal E}_{II}+\tilde{\mathcal E}_{I}\right)+2\tilde{H}_{1} \tilde{\mathcal E}_{KL}+ 32\pi^2 \tilde q_{(1)}\left(\tilde \mu+\tilde P+\frac{\tilde \Pi_{I}}{3}\right)
 +32\pi^2 \tilde q_{(2)}\tilde \Pi_{KL} ,\label{p1}
\end{eqnarray}
\end{widetext}
\begin{widetext}
\begin{eqnarray}
\tilde P_{(2)}=-\frac{2\tilde H_{1}}{3}\left(2 \tilde {\mathcal E}_{I}+\tilde {\mathcal E}_{II}\right)-2\tilde{H}_{2}\tilde {\mathcal E}_{KL}
+ 32\pi^2 \tilde q_{(2)}\left(\tilde \mu+\tilde P+\frac{\tilde \Pi_{II}}{3}\right) +32\pi^2\tilde q_{(1)}\tilde \Pi_{KL}. \label{SPP}
\end{eqnarray}
 \end{widetext}

In (\ref{p1}) and (\ref{SPP}), we can identify two different types of contributions. On the one hand, we have contributions from the  heat transport process. These are independent on the magnetic part of the Weyl tensor, and  appear in the tilted versions of LTB and Szekeres, as well as in the case analyzed in this~section.  

Next, we have contributions related to the gravitational radiation. These require both the electric and the magnetic part of the Weyl tensor to be different from zero. Of course, they vanish for LTB and Szekeres, but do not vanish in the present case.

The association of  a state of gravitational radiation to a  non-vanishing component of the  super-Poynting vector is enforced by the  link between the super-Poynting vector and the news functions in the context of the Bondi--Sachs approach \cite{5p}. 

Thus, we have in the case analyzed in this section that, for the comoving observer and the line element (\ref{len}), the magnetic part of the Weyl tensor vanishes identically and the fluid is non-dissipative, implying  at once that $\tilde P_{(1)} =\tilde P_{(2)}=0$. In other words, no gravitational radiation, or dissipative processes of any kind,  are detected by the comoving observer.

However, for the tilted congruence calculations show that the magnetic part of the Weyl tensor is not vanishing and, more specifically,  the sum of the first two terms in (\ref{p1}) and (\ref{SPP}) does not vanish, except for the conformally flat case \cite{hdc}. 

Thus, we face again the intriguing question: how it is possible that tilted observers may detect irreversible processes, whereas comoving observers describe an isentropic situation?

As we shall see,  the above quandary becomes intelligible if we appeal to the discussion on the Maxwell's demon presented by Bennet.

\section{THE MAXWELL'S DEMON AND TILTED AND COMOVING OBSERVERS}

The main moral emerging from the three cases analyzed here (and from many others included in the list of references) is that   tilted observers may detect dissipation in systems that appear non-dissipative for comoving observers. 

It is worth mentioning that, in the case analyzed in the previous section, the difference between the pictures described by both congruences of observers is still sharper since the tilted observer not only  detects a dissipative process, but also gravitational radiation. 

 This last point  is not alien to the fact  that   the tilted observer also detects vorticity, and as has been pointed out in \cite{5p}, vorticity  and gravitational radiation are tightly associated. At any rate, gravitational radiation is also a  dissipative process; accordingly, the basic explanation of its presence  in the system analyzed by the tilted observer is basically the same as the one  for any dissipative process.

As conjectured in \cite{en}, the basic fact that  explains the above-mentioned differences in the~description of  a given system, as provided by different congruences of observers, is  that  both congruences of observers store different amounts of information. 

 Here, we shall delve deeper into this question, by resorting to the resolution of the well known paradox of the Maxwell's demon \cite{max}. 

Let us first recall the Maxwell's demon paradox and how it was solved by Bennet, using the~theory of information. Let us take a look at Figure 1.
%
%
\begin{figure}
\includegraphics[width=3.5in,height=4.in,angle=0]{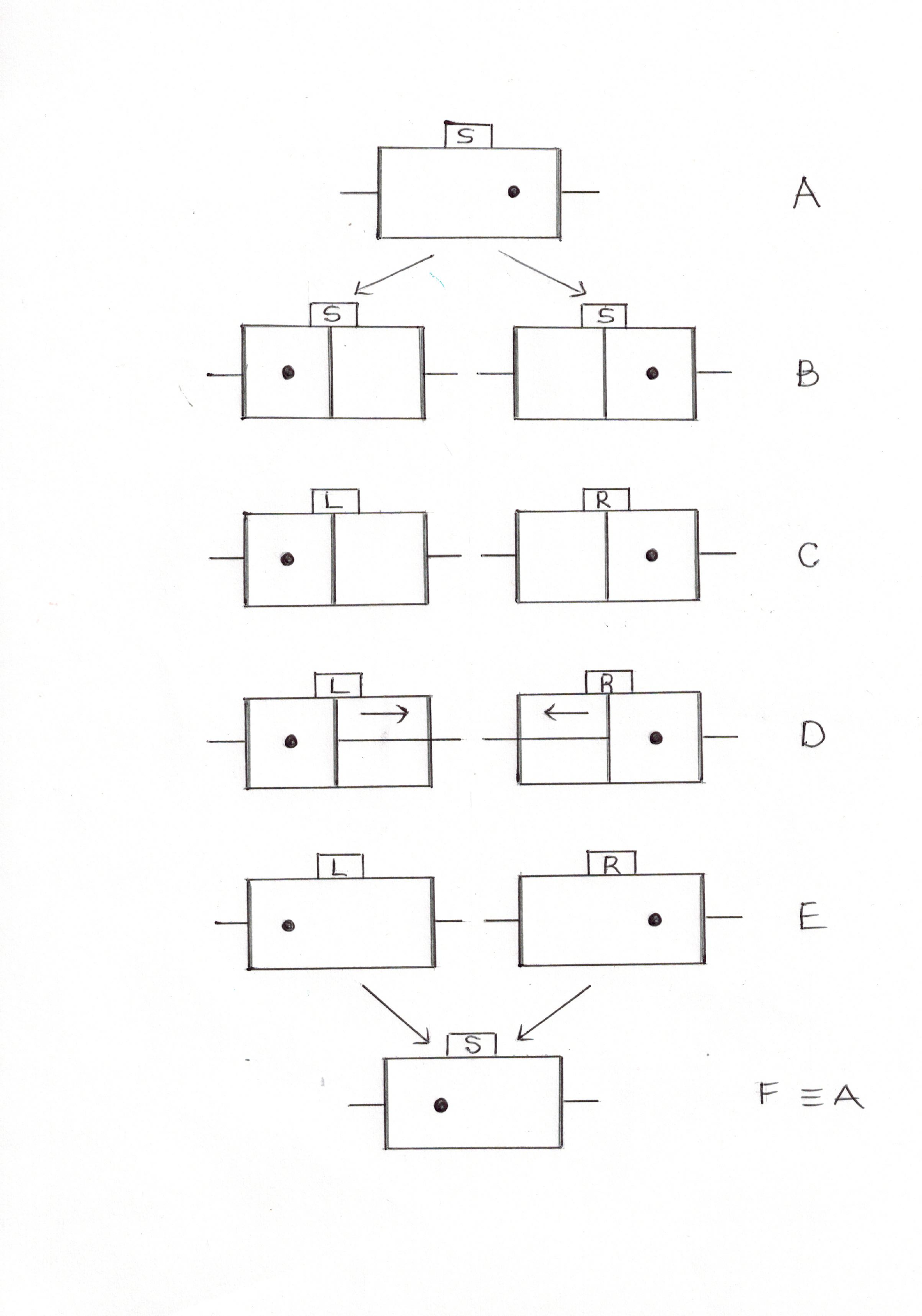}
\caption{The Bennet scheme.}
\label{fig:I}
\end{figure}


 Initially (stage A), we have a cylinder  containing one molecule with two pistons at either side. At this point, the demon  does not know where in the cylinder is the molecule. We shall refer to this state of the demon's mind as $S$.
 Next, in stage B, the demon inserts a partition wall in the middle of the~cylinder, trapping the molecule in one side or the other. In this stage, the demon still ignores on what side of the cylinder is the molecule; therefore, the state of his mind is still $S$.

 In stage C, the demon performs a reversible measurement allowing him to know whether the molecule is on the left or the right side of the cylinder. In each case, the state of the demon's mind changes to  $L$ or $R$, respectively.
 
In D, depending on the result of the previous measurement, the demon moves the left piston to the right (if the molecule is in the right), or the right piston to the left (if the molecule is in the left), and~removes the partition wall. Doing so, he allows the molecule to freely expand against the piston, and~thereby doing work.
 
In E, the pistons are in their original position and the molecule fills the whole cylinder. Thus, one is tempted to say that we have returned to the initial state A, but work has been done and therefore such a conclusion would imply the violation of the second law of thermodynamics.

 The solution to the above, apparent, paradox comes up when we realize that the demon's mind state in A and E are different. 
Indeed, in E, the demon knows where the molecule was before the expansion.
 In order to truly return to A, the information acquired by the demon has to be erased.
 
However, according to the Landauer principle \cite{lan}, the erasure of one bit of information stored in a system requires the dissipation into the environment of a minimal
amount of energy, whose lower bound is given by

\begin{equation}
\bigtriangleup E=kT \ln2,
\label{lan1}
\end{equation}
where $k$ and  $T$ denotes the Boltzman constant and the  temperature of the environment, respectively. 
 
In other words, to get the demon's mind back to  its initial state ($S$), generates dissipation, after which the system is in F.
 Thus, all the work obtained by the expansion of the molecule in D is converted to heat in order to return the demon's mind to the state $S$, in F.
 
Therefore, Bennet solved the paradox by showing that the irreversible act that prevents the violation of the second law is not the reversible measurement allowing him to know where the molecule is, but the resaturation of the measuring apparatus to the standard state prior to the state where the demon knows the location of the molecule. Therefore, if we consider the whole system (demon + the gas in the cylinder), we must keep in mind that the information possessed by  the demon, before  knowing the location of the molecule,  is smaller  than the information after this process has been achieved. Accordingly, in order to return to the initial state of the demon, the acquired information has to be~erased.

A somehow similar picture appears when we apply the operation transforming  comoving observers, who are assigned zero value to the three-velocity of any fluid element, into tilted observers, for whom the three-velocity represents another degree of freedom. The erasure of the information stored by comoving observers (vanishing three velocity), when going {in} the
frame of  tilted observers, must be accompanied by dissipation by virtue of the Landauer principle, which explains the presence of dissipative processes (included gravitational radiation) observed by the latter.

To check  the consistency of the explanation above, let us take a look at this issue  by considering the transition from the tilted congruence to the comoving one.

When passing from the tilted to the comoving congruence, a decrease of entropy occurs, but we do not have any external agent, and therefore such a decrease of entropy  is accounted for by the dissipative flux observed in the tilted congruence, which leads to a isentropic system, as seen by the comoving congruence (we recall  that, in the comoving congruence, the system is dissipationless). In~other words, all the dissipation detected by the tilted congruence is associated with the information difference between both congruences.

Thus, we can say that the state $S$ of the demon, when he does not know the location of the molecule, is analogous  to tilted observers:  for both, a piece of information is lacking. On the other hand, the state $L$ or $R$ when the demon knows the location of the molecule, is equivalent to  comoving observers: in both cases,  additional information has been  acquired.

\section{Discussion}

With the three examples analyzed in the previous sections, we have clearly illustrated the relevance  of  observers in the physical description of a given system. 

To explain the detection  of dissipation by the tilted congruence, in a system which appears  isentropic for comoving observers, we have noticed that passing from comoving to tilted observers, or returning the demon's mind to its initial state, requires the erasure of the acquired information, leading to the observed dissipative processes. This explains, on the one hand,  why the second law of thermodynamics is not violated by the Maxwell's demon, and, on the other hand, why tilted observers detect dissipation there when comoving observers only see an isentropic system.

In other words, observers storing different amounts of information provide different pictures of the same phenomenon.

In light  of the comments above, the statement by Max Born  \cite{mb} {\it ``Irreversibility is a consequence of the explicit introduction of ignorance into the fundamental laws''}
 becomes fully intelligible.

\section*{Acknowledgments}
{This  work  was partially supported by the Spanish  Ministerio de Ciencia e
Innovaci\'on under Research Projects No.  FIS2015-65140-P (MINECO/FEDER)}.

\end{document}